\documentclass[conference]{IEEEtran}
\IEEEoverridecommandlockouts
\usepackage{cite}
\usepackage{amsmath,amssymb,amsfonts}
\usepackage{algorithmic}
\usepackage{graphicx}
\usepackage{textcomp}
\usepackage{xcolor}
\usepackage{subfig}
\usepackage{siunitx, xfrac}
\usepackage{glossaries}
\usepackage{url} 
\usepackage[hidelinks]{hyperref}

\renewcommand{\url}[1]{\nolinkurl{#1}}
\hypersetup{draft}

\def\BibTeX{{\rm B\kern-.05em{\sc i\kern-.025em b}\kern-.08em
    T\kern-.1667em\lower.7ex\hbox{E}\kern-.125emX}}

\newacronym{XR}{XR}{Extended Reality}
\newacronym{ISAC}{ISAC}{Integrated Sensing and Communications}
\newacronym{DL}{DL}{Deep Learning}
\newacronym{CNN}{CNN}{Convolutional Neural Network}
\newacronym{mmWave}{mmWave}{Millimeter-Wave}
\newacronym{COTS}{COTS}{Commercial Off-The-Shelf}
\newacronym{OFDM}{OFDM}{Orthogonal Frequency-Division Multiplexing}
\newacronym{SNR}{SNR}{Signal-to-Noise Ratio}

\graphicspath{{figures/}}

\begin{document}

\title{Millimeter-Wave Gesture Recognition in ISAC:\\Does Reducing Sensing Airtime Hamper Accuracy?}

\author{
\IEEEauthorblockN{Jakob Struye\IEEEauthorrefmark{1}, Nabeel Nisar Bhat\IEEEauthorrefmark{1}, Siddhartha Kumar\IEEEauthorrefmark{2}, Mohammad Hossein Moghaddam\IEEEauthorrefmark{2}, Jeroen Famaey\IEEEauthorrefmark{1}\\
\IEEEauthorblockA{\IEEEauthorrefmark{1}University of Antwerp - imec, Antwerp, Belgium. Email: \{firstname\}.\{lastname\}@uantwerpen.be }
\IEEEauthorblockA{\IEEEauthorrefmark{2} Qamcom Research and Technology AB, Gothenburg, Sweden. Email: \{siddhartha.kumar,mh.moghaddam\}@qamcom.se}
}}

\maketitle

\begin{abstract}
Most Integrated Sensing and Communications (ISAC) systems require dividing airtime across their two modes. However, the specific impact of this decision on sensing performance remains unclear and underexplored. In this paper, we therefore investigate the impact on a gesture recognition system using a Millimeter-Wave (mmWave) ISAC system. With our dataset of power per beam pair gathered with two mmWave devices performing constant beam sweeps while test subjects performed distinct gestures, we train a gesture classifier using Convolutional Neural Networks. We then subsample these measurements, emulating reduced sensing airtime, showing that a sensing airtime of \SI{25}{\percent} only reduces classification accuracy by \num{0.15} percentage points from full-time sensing. Alongside this high-quality sensing at low airtime, mmWave systems are known to provide extremely high data throughputs, making mmWave ISAC a prime enabler for applications such as truly wireless Extended Reality.
\end{abstract}

\section{Introduction}
Within the \gls{ISAC} paradigm, the same set of hardware offers both sensing and communications capabilities within some system, which is expected to have an impact on the performance of both modes. Some recent works have investigated this tradeoff from a theoretical perspective in a 6G context, reusing the same \gls{OFDM} waveform for the two modes~\cite{tradeoff4,tradeoff2,tradeoff3,tradeoff1}. In this work, we take a different approach, considering lower-cost in-home \gls{ISAC} systems, where hardware is expected to alternate between the two modes~\cite{ericsson}, and evaluate the tradeoff from an application perspective. For communications, the impact is straightforward, with throughput being directly proportional to communications airtime. For the sensing mode, however, the impact is more difficult to gauge, and is currently poorly understood. Intuitively, the accuracy of sensing results is expected to drop as the sensing airtime is reduced, with the exact impact likely depending on the specific type of sensing. As such, we investigate the magnitude of this impact specifically for gesture recognition within this paper. 

Several applications of \gls{ISAC} for gesture recognition exist within the home environment. Gesture recognition can replace television remotes~\cite{remote}, serve as a more intuitive input for \gls{XR} applications~\cite{xr,towardsxr}, or be used to control home automation features~\cite{homeautomation}. Compared to traditional camera-based systems, \gls{ISAC} not only alleviates privacy concerns, but also provides a significant cost reduction. One dedicated device, or even existing home Wi-Fi infrastructure, can provide sensing capabilities alongside any wireless communications necessary. Most research focuses on the commonly used 2.4 and \SI{5}{\giga\hertz} Wi-Fi bands, proving the viability of this approach~\cite{wifisurvey,wigrunt,wiopen}. However, the \gls{mmWave} band, between 30 and \SI{300}{\giga\hertz}, may offer more accurate gesture recognition through its shorter wavelength and larger antenna arrays, which enable a higher sensing resolution. 
\begin{figure*}[!t]
\centering
\includegraphics[width=\linewidth]{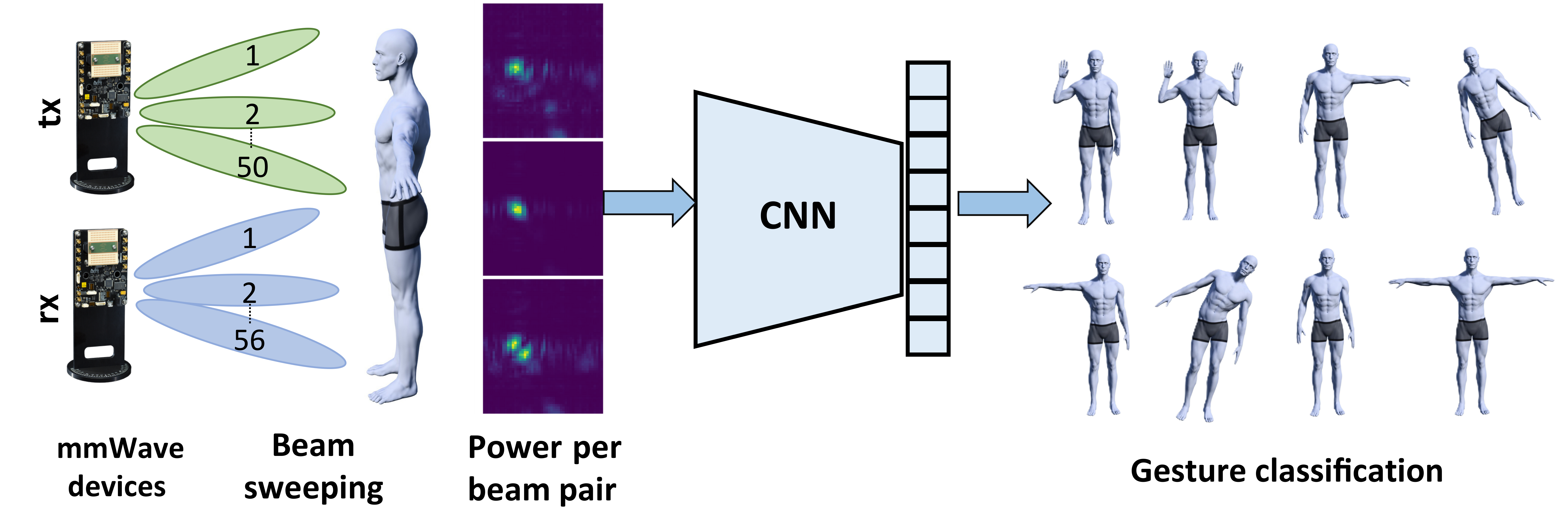}
\caption{The end-to-end system, going from \gls{mmWave} power per beam pair measurements to gesture classification.}
\label{fig:pipeline}
\end{figure*}

Initial work has shown the advantages of \gls{mmWave} with \gls{COTS} devices. Yu \textit{et al.} leverage \gls{COTS} mmWave Wi-Fi devices for human pose detection~\cite{gestureset1}. A \gls{DL}-based system is used to map changes in beam \glspl{SNR} to different poses, achieving \SI{88}{\percent} classification accuracy. Bhat \textit{et al.} use mmWave \gls{COTS} Wi-Fi for gesture recognition~\cite{bhat2023gesture}, with a \gls{DL}-based classifier achieving \SI{96.7}{\percent} accuracy in a single-person, single-environment setting, and \SI{87}{\percent} accuracy across multiple environments and diverse users. Wang \textit{et al.} investigate the feasibility of gesture recognition using \gls{mmWave} Wi-Fi's beam sweeping protocol~\cite{slssense}. Furthermore, \gls{COTS} \gls{mmWave} has shown promise in related fields such as skeletal pose estimation~\cite{gestureset2}, trajectory estimation~\cite{trajectory}, localization and tracking~\cite{loctrack}, vital sign detection~\cite{vitalsigns} and violent activity detection~\cite{violent}. 

In contrast to these works, which focus on \gls{COTS} devices, we consider a dataset with state-of-the-art experimental \gls{mmWave} hardware, providing full control to the experimenter. The measurements in this dataset leverage \textit{beamforming}, a core feature in \gls{mmWave} networking, in which both transmission and reception are focused in some direction, through an \textit{antenna array} of many small antenna elements~\cite{mmWaveBook}. Compared to lower-frequency communications, beamforming with \gls{mmWave} is both more necessary, as signals fade more rapidly, and more feasible, as more of its small antenna elements can be packed together. Fig.~\ref{fig:pipeline} summarizes our system, in which rapid \textit{beam sweeps} gathered with 50 transmit beams and 56 receive beams are fed to a \gls{DL}-based gesture classifier.

In the above dataset, \SI{100}{\percent} of airtime was dedicated to sensing, which clearly would not classify as a true \gls{ISAC} system. As such, we investigate the impact on classification performance from reducing this sensing airtime, freeing up airtime for communications. We achieve sensing airtime reduction by either performing \textit{fewer} or \textit{shorter} beam sweeps, and investigate how the approach taken impacts final performance. In addition, we investigate methods towards finding the optimal airtime reduction approach.

\section{Dataset}
A total of 7 test subjects participated in our measurement campaign. They each performed a set of 8 gestures, inspired by prior works~\cite{gestureset1,gestureset2}, including both static (e.g., leaning) and dynamic (e.g., moving arms up and down) as shown in Fig.~\ref{fig:pipeline}. Each test subject performed 7 full sequences of all gestures, maintaining each gesture for \SI{10}{\second} at a time, meaning each test subject performed gestures for \SI{560}{\second} in total. 

\begin{figure*}[!t]
\centering
\includegraphics[width=\linewidth]{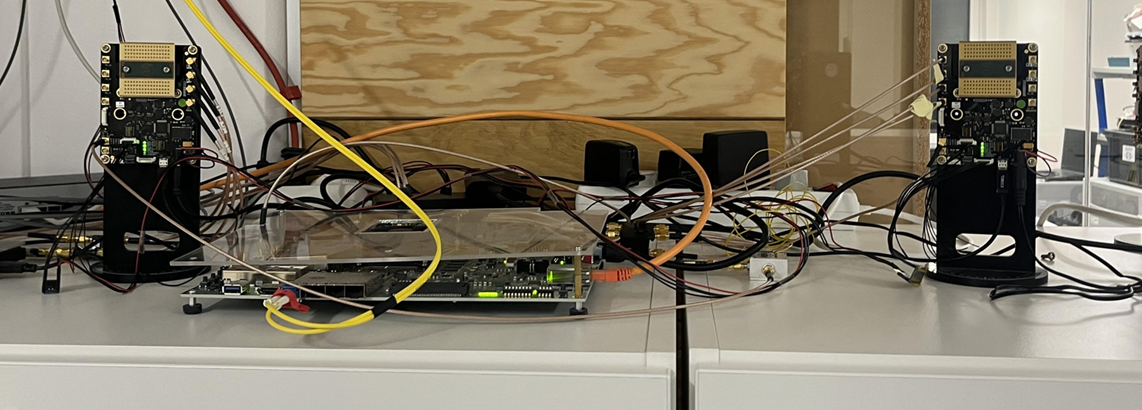}
\caption{The experimental setup, as seen from the test subject's point of view. The transmitting and receiving devices are placed side-by-side and both aimed towards the test subject.}
\label{fig:setup}
\end{figure*}

For sensing, we employed two \gls{mmWave} Sivers Evaluation Kit EVK06002~\cite{sivers}. The devices support the 57-\SI{71}{\giga\hertz} range, and were set to a \SI{760}{\mega\hertz} bandwidth centered around \SI{60}{\giga\hertz} for these experiments. They were placed next to each other with their broadsides aimed towards the user, such that the receiving device would mostly receive the transmitter's signal through reflections off the test subject's body, as shown in Fig.~\ref{fig:setup}. The two devices coordinated to rapidly cycle through their 50 transmit beams and 56 receive beams, resulting in \num{2800} possible beam pairs. The devices performed \num{154} full beam sweeps per second, resulting in  each beam pair being active for \SI{2.319}{\micro\second} at a time. During each such interval, exactly two \gls{OFDM} symbols are transmitted. The final measured power per beam pair is obtained by convolving the raw transmit and receive signals.

\section{Gesture Recognition System}
In our dataset, \SI{100}{\percent} of airtime was dedicated to beam sweeping for sensing. Realistically, airtime would be divided between communications and sensing. While the performance impact of reducing communications airtime in terms of throughput and latency is rather straightforward to gauge, this is not the case for the sensing aspect, especially when a \gls{DL} approach is taken to distill raw measurements into practical sensing results, as shown in Fig.~\ref{fig:pipeline}. As such, we leverage the measurements from our dataset to quantify the impact of reduced sensing airtime on sensing accuracy. We can artificially reduce the airtime in the dataset by simply subsampling the full data. We then train a \gls{CNN}-based deep neural network as a gesture classifier, trained with different levels of sensing airtime. 

\subsection{Preprocessing}
For these experiments, each beam sweep was preprocessed to a 2D array of power measurements, and a sliding window of the 20 most recent beam sweeps is used as input for the classifier. This results in an input array of dimensions $(t,tx,rx)$ where $t=20$, $tx=50$ and $rx=56$. To emulate reduced sensing airtime, we can subsample along any of these axes, and these experiments will show how sensitive the classifier is to this selection. Specifically, we can emulate either performing fewer beam sweeps (subsampling \textit{in time}, by reducing $t$) or performing shorter beam sweeps (subsampling \textit{in space}, by reducing $tx$ and/or $rx$). In each case, subsampling with factor $s$ is implemented by only keeping every $s^{\textnormal{th}}$ measurement along the axis, starting with the first. For each single-axis subsampling, we experiment with $s=2,3,\dots,9$ (i.e., \SI{50}{\percent} to $\sim$\SI{11}{\percent} sensing airtime). When subsampling both $tx$ and $rx$, we consider $s=\{4,9\}$, subsampling with factor $\sqrt{s}$ across each axis. For the remainder of this paper, we will use `subsampling factor' and `sensing airtime fraction' interchangeably.

\subsection{Classifier Architecture}
We use the same, relatively standard, network architecture for all experiments and forego any extensive hyperparameter tuning, as we are investigating the relative performance given different airtimes, rather than maximizing classification performance. We opt for a three-layer \gls{CNN} where each convolutional block consists of a 2D convolutional layer, batch normalization and a ReLU activation function. The first two blocks use $3 \times 3$ kernels with 16 and 32 output channels respectively, while the third has a $7 \times 7$ kernel and $64$ output channels. Each block is followed by a $2 \times 2$ max pooling layer. This output is flattened and passed to a fully connected layer with 8 outputs for final classification. This is trained using the Adam optimizer with cross-entropy loss. During inference, the class with the highest output is chosen.

\subsection{Upsampling}
At high subsampling factors, the input dimensions become smaller than the minimum size required for the \gls{CNN} above. As such, we apply an upsampling step to scale the axes of the subsampled data back up to the original dimensions. We achieve this by simply repeating each measurement $s$ times in a row, truncating the axis at the end when the original dimension was not a multiple of $s$. As a simple one-dimensional example, subsampling the array $[v_0,v_1,v_2,v_3,v_4]$ with $s=3$ leads to $[v_0,v_0,v_0,v_3,v_3]$. We also experimented with repeating the \textit{full} subsampled array $s$ times ($[v_0,v_3,v_0,v_3,v_0]$), but initial experimentation showed this had a significant negative impact on classification accuracy. This result was expected, as adjacent beam indices imply similar beam directions, and \glspl{CNN}, by design, exploit such spatial locality.

\subsection{Training and Testing}
With 8 different values for $s$ for each of the three axes with single-axis subsampling, and 2 values for $tx+rx$ subsampling, along a baseline without subsampling, we train and evaluate the classifier for 27 different variants of the dataset. In each case, the dataset is split into train, validation and test sets with a $(72,8,20)$ split. Here, we are careful to take the \textit{first} \SI{72}{\percent} of measurements for each combination of test subject and performed gesture as training data, followed by the next \SI{8}{\percent} and finally \SI{20}{\percent}. By carefully splitting with this approach, and only shuffling data afterwards, we avoid any \textit{information bleeding} between the different sets, as a temporal sliding window was used during initial parsing. Through initial informal hyperparameter tuning, we decided to perform $100$ epochs of training with learning rate $3\text{e-}4$ and batch size $512$, offering an acceptable tradeoff between training time and accuracy. As final classification accuracy, we take the performance on the test set for the result of the epoch that achieved the highest accuracy on the validation set. We repeat each training 25 times, reporting the mean and standard deviation of the classification accuracy. To reduce wall-clock runtime, these experiments were run on an internal cluster, using a mix of NVIDIA Quadro RTX 4000 and Tesla V100 GPUs, with each epoch requiring approximately \SI{5}{\second} of compute time on one GPU. We provide the full experimental code on GitHub\footnote{https://github.com/JakobStruye/isac-tradeoff}.
\begin{figure*}[!t]
\centering
\includegraphics[width=\linewidth]{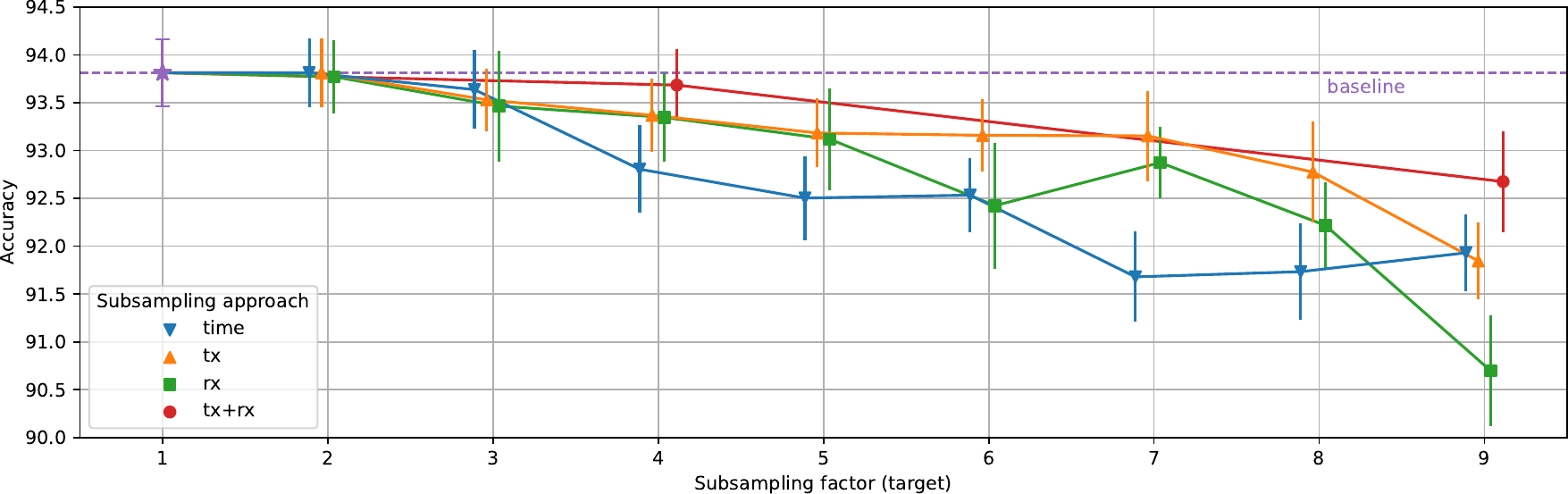}
\caption{Classification accuracy for different subsampling factors after 100 epochs. For each datapoint, training was repeated 25 times, with mean and $\pm$1 standard deviation visualized. Slight horizontal offsets are introduced during visualization to make error bars more distinguishable (all correct values for the subsampling factor are integers). Subsampling factor 1 reverts to the baseline (\SI{100}{\percent} sensing airtime) for each approach, indicated with a purple star and horizontal dashed line at baseline performance level.}
\label{fig:performance_100}
\end{figure*}

\section{Results}
To investigate the impact of reducing the relative sensing airtime on the quality of sensing results, we primarily investigate the accuracy of the gesture classifier for different sensing airtimes, emulated by subsampling the available measurements. 
\subsection{Sensing airtime impact}
\begin{figure*}[!t]
\centering
\subfloat[Baseline (\SI{100}{\percent} sensing airtime)\label{fig:confmat_base}]{%
    \includegraphics[width=0.4\textwidth]{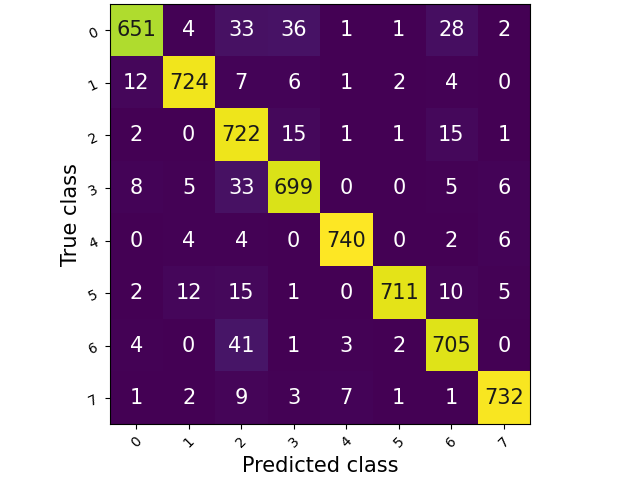}
}\hskip1ex
\subfloat[Subsampled with factor 9 across tx and rx\label{fig:confmat_optim}]{%
    \includegraphics[width=0.4\textwidth]{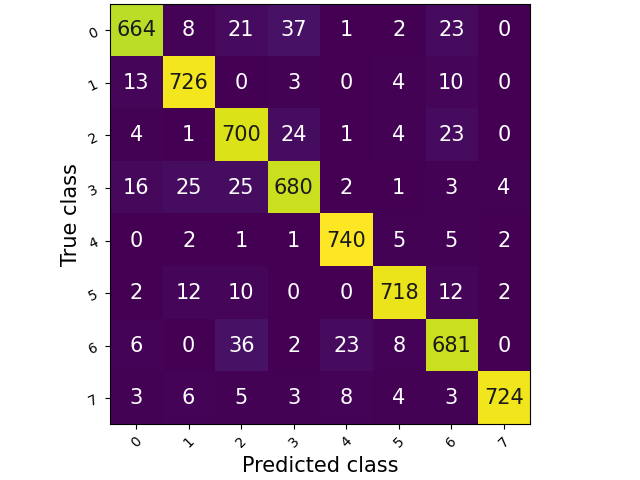}
}
\caption{Two arbitrarily selected confusion matrices showing that, even with severe subsampling, the classifier performs well for all classes.}
\label{fig:confmat}
\end{figure*}
Fig.~\ref{fig:performance_100} shows these results for a classifier trained for 100 epochs. For each subsampling factor $s$ and subsampling approach, training was repeated 25 times, from which mean $\mu$ and standard deviation $\sigma$ were derived. The figure shows $\mu$ as datapoint with error bars indicating $[\mu-\sigma,\mu+\sigma]$. With each approach, $s=1$ reduces to the baseline, shown with a star. The figure shows that subsampling in time leads to a more rapid reduction in performance, losing on average more than 1 percentage point at $s=4$, with each in-space subsampling approach losing less than 0.5 percentage points at $s=4$ (a \SI{75}{\percent} reduction in sensing airtime). For higher $s$, the different in-space subsampling approaches begin to deviate from each other, with rx-only performing the worst, and tx+rx subsampling being the best approach, experiencing an accuracy reduction of only 1.15 percentage points for $s=9$. In-time subsampling appears to eventually overtake in-space, but discussion on this is deferred to Sec.~\ref{sec:actualtime}. When inspecting the confusion matrices for the different airtimes, with some shown in Fig.~\ref{fig:confmat}, some classes, most notably the first, appear more difficult to predict correctly. As this behavior remains consistent with reduced airtime, we consider any further investigation of this outside of the scope of this work. Overall, performance decreases from airtime reduction do not appear to impact any classes excessively.

\subsection{Sensitivity to training time}
One potential concern with these results is the possibility that training converges more rapidly with fewer distinct measurements, such that the classifiers relying on a higher subsampling factor are more ``fully'' trained than those with lower subsampling factors. To investigate this, we simply repeat the entire test suite with double the amount of epochs, again with 25 repeats. The mean accuracy for each scenario increases by 0.4 to 0.9 percentage points, and there is no discernible pattern of higher increases with more subsampling. From this, we conclude that training ``progress'' was similar between all scenarios. We note that, as we are investigating \textit{relative} performance, we consider epoch hyperparameter tuning as outside the scope of this work.
\begin{figure*}[!t]
\centering
\includegraphics[width=\linewidth]{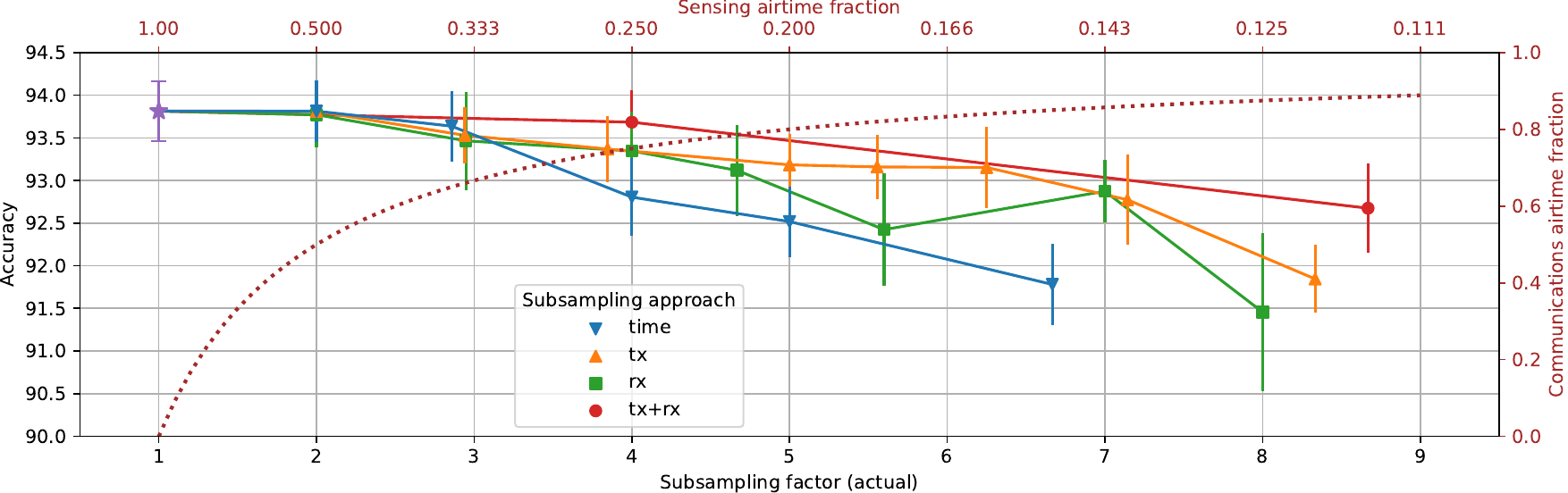}
\caption{The classification accuracy results from Fig.~\ref{fig:performance_100} re-plotted with x-axis values adjusted to show the \textit{actual} rather than \textit{target} subsampling factor (now without horizontal offset) after 100 epochs. The brown line and axes show the airtime fractions for sensing and communications. The communications axis is kept linear to emphasize how rapidly it increases with increasing sensing subsampling. The horizontal baseline performance line is removed for visual clarity.}
\label{fig:performance_100_actual}
\end{figure*}

\subsection{Actual vs target subsampling factors}\label{sec:actualtime}
An important observation on the subsampling factors is that it can only be achieved exactly when $s$ is a divisor of $d$, the length of the axis being subsampled on. With our approach, the \textit{actual} subsampling factor $s'$ will be lower than the \textit{target} subsampling factor $s$, adhering to the equation $s = d\lceil\frac{d}{s}\rceil^{-1}$, which simplifies to $s = s'$ i.f.f. $d$ is divisible by $s$. Intuitively, it is obvious that the difference between $s$ and $s'$ can be larger the smaller $d$ is. In some cases, different values of $s$ may even lead to the \textit{same} $s'$. This is most pronounced for $d=20$ (i.e., time-based subsampling), where $s=5$ and $s=6$ both lead to $s'=5$, and target subsampling factors 7, 8 and 9 all lead to an actual factor of $6.67$. We do note that, while the number of selected values in the subsampling is equal in these cases, the exact selection itself will differ for different values of $s$. To get a more complete view of the impact of subsampling, we re-plot Fig.~\ref{fig:performance_100} with the actual subsampling factor on the x-axis, resulting in Fig.~\ref{fig:performance_100_actual}. If multiple values of $s$ map to the same value $s'$, all of these results were used to calculate the mean and standard deviation. Generally, findings that were based on the target subsampling factor still hold with the actual subsampling factor. The one exception is that, with actual subsampling factors, in-time subsampling performs consistently worse than in-space, even for higher $s'$ (as the actual factor is significantly lower than the target one).

As an addition, this plot contains a second x-axis, showing the airtime fraction used for sensing. The axis serves as an alternative x-axis for the accuracy plots (i.e., it lines up with the primary x-axis). Using a secondary y-axis, the dashed brown line shows the relationship between sensing airtime (or subsampling factor) and communications airtime fractions. Unlike the secondary x-axis, this y-axis is kept linear, to more intuitively show how rapidly the communications airtime increases with the subsampling factor. 
\begin{figure}[!t]
\centering
\includegraphics[width=\linewidth]{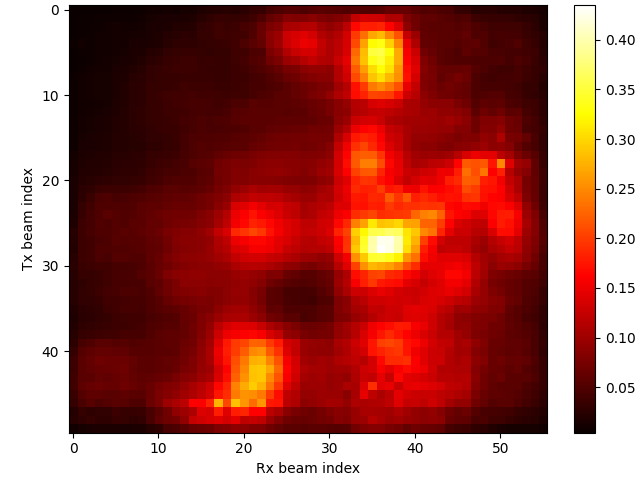}
\caption{Spatial saliency map derived from a classifier trained on the non-subsampled dataset}
\label{fig:heatmap}
\end{figure}

\subsection{Beam pair selection}
These results convincingly show that subsampling in space (i.e., beam pairs) is preferable to subsampling in time when optimizing for gesture classification accuracy. Furthermore, splitting the subsampling across tx and rx beams performs better than focusing on one of the two for a given subsampling factor. Also, the exact selection of beam pairs can have a tangible impact, as illustrated by the accuracy increasing when going from target rx subsampling factor 6 to 7 in Fig.~\ref{fig:performance_100}. As such, for some target subsampling factor, there must be an optimal selection of beam pairs, and a poor selection could have a noticeable impact on performance. One approach to intelligently select these beams is to first generate a saliency map of the beam pairs~\cite{saliency}. This two-dimensional ($50 \times 56$) saliency map shows how strongly a small change in a beam pair's measurement would impact the classification. We generate this by running a random selection of 1000 test samples through a trained predictor and, for each sample, calculating the (absolute) gradients through backpropagation. As we are only interested in beam pair saliency, the results are averaged over the temporal axis. Fig.~\ref{fig:heatmap} shows a saliency map for the baseline, non-subsampled classifier. Some regions have a significant impact on classification, while the result is mostly insensitive to other regions. This makes sense, as some beams physically miss the test subject entirely. Clearly, such a saliency can aid in intelligently subsampling beam pairs. However, this process is not straightforward. The saliency map is expected to be dynamic over time, meaning that deriving it would require regularly re-training a non-subsampled model, which is resource-intensive and requires full beam sweeps in a live system which is otherwise subsampled, temporarily reducing communications airtime. Then, each time the beam pair selection changes, the actual classifier needs to be retrained (or at least finetuned). Furthermore, it is not immediately clear how the selected beam pairs should be arranged into a grid. As adjacent beam indices imply similar beam directions, the \gls{CNN} leverages spatial information from the grid layout, which is difficult to maintain with ``non-uniform'' subsampling. To corroborate this, we re-trained the baseline classifier with arbitrarily re-assigned beam indices, leading to an accuracy reduction of approximately 2 percentage points. Overall, given these difficulties, alongside the already minor performance gap between non-subsampled and uniformly subsampled beam pairs, we do not investigate this any further. 

\section{Conclusions}
In this paper, we presented a novel dataset for gesture classification in a \gls{mmWave} \gls{ISAC} system. For 7 test subjects each performing 8 distinct gestures sequentially, two \gls{mmWave} devices rapidly cycled through all \num{2800} transmit/receive beam pair combinations while measuring power. We then investigate how well a \gls{CNN}-based classifier can determine the performed gesture based on these \gls{mmWave} measurements, achieving \SI{93.8}{\percent} accuracy after 100 epochs. We performed this for multiple ways of subsampling the measurements, emulating an environment in which the \gls{ISAC} system only dedicates some fraction of airtime to sensing. When subsampling uniformly in both transmit and receive beams, the sensing airtime can be reduced to around \SI{25}{\percent} resulting in an accuracy reduction of only 0.15 percentage points. A different subsampling approach or higher subsampling factor has a significantly worse effect on performance, showing an informed selection of the two is important. Overall, these results show that \gls{mmWave} is a promising enabler for affordable in-home \gls{ISAC} systems which need to provide both extreme throughput and accurate sensing, such as for \gls{XR} applications~\cite{towardsxr}.
\section*{Acknowledgments}
This research was partially funded by the Research Foundation - Flanders (FWO) project WaveVR (Grant number G034322N). This work is partially supported by the European Commission through the Horizon Europe JU SNS project Hexa-X-II (Grant Agreement no. 101095759). Nabeel Nisar Bhat is supported by an FWO SB PhD fellowship (Grant number 1SH5X24N).
 
%

\bibliographystyle{IEEEtran}
\bibliography{IEEEabrv,bibliography}

@INPROCEEDINGS{gestureset1,
  author={Yu, Jianyuan and Wang, Pu and Koike-Akino, Toshiaki and Wang, Ye and Orlik, Philip V. and Sun, Haijian},
  booktitle={2020 IEEE Globecom Workshops}, 
  title={Human Pose and Seat Occupancy Classification with Commercial MMWave WiFi}, 
  year={2020},
  volume={},
  number={},
  pages={1-6},
  doi={10.1109/GCWkshps50303.2020.9367535}}

@inproceedings{gestureset2,
author = {Bhat, Nabeel Nisar and Sameri, Javad and Struye, Jakob and Vega, Maria Torres and Berkvens, Rafael and Famaey, Jeroen},
title = {Multi-modal pose estimation in XR applications leveraging integrated sensing and communication},
year = {2023},
doi = {10.1145/3615452.3617944},
booktitle = {Proceedings of the 1st ACM Workshop on Mobile Immersive Computing, Networking, and Systems},
pages = {261–267},
numpages = {7},
location = {Madrid, Spain},
}

@book{MmWaveBook,
  title={Millimeter wave wireless communications},
  author={Rappaport, Theodore S and Heath Jr, Robert W and Daniels, Robert C and Murdock, James N},
  year={2015},
  publisher={Pearson Education}
}

@INPROCEEDINGS{homeautomation,
  author={Zou, Han and Zhou, Yuxun and Yang, Jianfei and Jiang, Hao and Xie, Lihua and Spanos, Costas J.},
  booktitle={2018 IEEE 14th International Conference on Control and Automation (ICCA)}, 
  title={WiFi-enabled Device-free Gesture Recognition for Smart Home Automation}, 
  year={2018},
  volume={},
  number={},
  pages={476-481},
  doi={10.1109/ICCA.2018.8444331}}

@inproceedings{xr,
author = {Chen, Yi-Jing and Huang, Huai-Sheng},
title = {Gesture Recognition applied to Extended Reality: A Case Study of Online Meeting},
year = {2024},
doi = {10.1145/3657547.3657551},
booktitle = {Proceedings of the 2024 8th International Conference on Virtual and Augmented Reality Simulations},
pages = {84–89},
numpages = {6},
}

@ARTICLE{towardsxr,
  author={Struye, Jakob and Van Damme, Sam and Bhat, Nabeel Nisar and Troch, Arno and Van Liempd, Barend and Assasa, Hany and Lemic, Filip and Famaey, Jeroen and Vega, Maria Torres},
  journal={IEEE Communications Magazine}, 
  title={Toward Interactive Multi-User Extended Reality Using Millimeter-Wave Networking}, 
  year={2024},
  volume={62},
  number={8},
  pages={54-60},
  doi={10.1109/MCOM.001.2300804}}

@article{remote,
author = {Huiyue Wu and Liuqingqing Yang and Shengqian Fu and Xiaolong (Luke) Zhang},
title ={Beyond remote control: Exploring natural gesture inputs for smart TV systems},

journal = {Journal of Ambient Intelligence and Smart Environments},
volume = {11},
number = {4},
pages = {335-354},
year = {2019},
doi = {10.3233/AIS-190528},
}

@inproceedings{bhat2023gesture,
  title={Gesture recognition with mmWave Wi-Fi access points: Lessons learned},
  author={Bhat, Nabeel Nisar and Berkvens, Rafael and Famaey, Jeroen},
  booktitle={2023 IEEE 24th International Symposium on a World of Wireless, Mobile and Multimedia Networks (WoWMoM)},
  pages={127--136},
  year={2023},
  organization={IEEE}
}

@article{wifisurvey,
author = {Ma, Yongsen and Zhou, Gang and Wang, Shuangquan},
title = {WiFi Sensing with Channel State Information: A Survey},
year = {2019},
issue_date = {May 2020},
volume = {52},
number = {3},
issn = {0360-0300},
doi = {10.1145/3310194},
journal = {ACM Computing Surveys},
articleno = {46},
numpages = {36},
}

@ARTICLE{wigrunt,
  author={Gu, Yu and Zhang, Xiang and Wang, Yantong and Wang, Meng and Yan, Huan and Ji, Yusheng and Liu, Zhi and Li, Jianhua and Dong, Mianxiong},
  journal={IEEE Transactions on Human-Machine Systems}, 
  title={WiGRUNT: WiFi-Enabled Gesture Recognition Using Dual-Attention Network}, 
  year={2022},
  volume={52},
  number={4},
  pages={736-746},
  doi={10.1109/THMS.2022.3163189}}

@ARTICLE{wiopen,
  author={Zhang, Xiang and Huang, Jinyang and Yan, Huan and Feng, Yuanhao and Zhao, Peng and Zhuang, Guohang and Liu, Zhi and Liu, Bin},
  journal={IEEE Transactions on Human-Machine Systems}, 
  title={WiOpen: A Robust Wi-Fi-Based Open-Set Gesture Recognition Framework}, 
  year={2025},
  volume={55},
  number={2},
  pages={234-245},
  doi={10.1109/THMS.2025.3532910}}

@ARTICLE{slssense,
  author={Wang, Jian and Chuang, Jack and Berweger, Samuel and Gentile, Camillo and Golmie, Nada},
  journal={IEEE Internet of Things Journal}, 
  title={Toward Opportunistic Radar Sensing Using Millimeter-Wave Wi-Fi}, 
  year={2024},
  volume={11},
  number={1},
  pages={188-200},
  doi={10.1109/JIOT.2023.3301006}}

@INPROCEEDINGS{trajectory,
  author={Vaca-Rubio, Cristian J. and Wang, Pu and Koike-Akino, Toshiaki and Wang, Ye and Boufounos, Petros and Popovski, Petar},
  booktitle={ICASSP 2023 - 2023 IEEE International Conference on Acoustics, Speech and Signal Processing}, 
  title={mmWave Wi-Fi Trajectory Estimation with Continuous-Time Neural Dynamic Learning}, 
  year={2023},
  volume={},
  number={},
  pages={1-5},
  doi={10.1109/ICASSP49357.2023.10096474}}

@INPROCEEDINGS{loctrack,
  author={Wang, Jian and Chuang, Jack and Semper, Sebastian and Golmie, Nada},
  booktitle={2024 33rd International Conference on Computer Communications and Networks (ICCCN)}, 
  title={Super-resolution Localization and Tracking in WiFi Sensing}, 
  year={2024},
  volume={},
  number={},
  pages={1-9},
  doi={10.1109/ICCCN61486.2024.10637535}}

@INPROCEEDINGS{vitalsigns,
  author={Blandino, Steve and Bang, Jihoon and Wang, Jian and Berweger, Samuel and Chuang, Jack and Senic, Jelena and Ropitault, Tanguy and Gentile, Camillo and Golmie, Nada},
  booktitle={ICASSP 2024 - 2024 IEEE International Conference on Acoustics, Speech and Signal Processing}, 
  title={Low Overhead DMG Sensing for Vital Signs Detection}, 
  year={2024},
  volume={},
  number={},
  pages={13041-13045},
  doi={10.1109/ICASSP48485.2024.10446367}}

@INPROCEEDINGS{violent,
  author={P, Sruthi. and Udgata, Siba K},
  booktitle={2023 IEEE 11th Region 10 Humanitarian Technology Conference (R10-HTC)}, 
  title={Wi-Fi Sensing Enabled Violent Activity Detection in a Smart Home}, 
  year={2023},
  volume={},
  number={},
  pages={1274-1279},
  doi={10.1109/R10-HTC57504.2023.10505060}}

@inproceedings{saliency,
author = {Alqaraawi, Ahmed and Schuessler, Martin and Wei\ss{}, Philipp and Costanza, Enrico and Berthouze, Nadia},
title = {Evaluating saliency map explanations for convolutional neural networks: a user study},
year = {2020},
isbn = {9781450371186},
doi = {10.1145/3377325.3377519},
booktitle = {Proceedings of the 25th International Conference on Intelligent User Interfaces},
pages = {275–285},
numpages = {11},
location = {Cagliari, Italy},
}

@INPROCEEDINGS{tradeoff1,
  author={Xiong, Yifeng and Liu, Fan and Cui, Yuanhao and Yuan, Weijie and Han, Tony Xiao},
  booktitle={GLOBECOM 2022 - 2022 IEEE Global Communications Conference}, 
  title={Flowing the Information from Shannon to Fisher: Towards the Fundamental Tradeoff in ISAC}, 
  year={2022},
  volume={},
  number={},
  pages={5601-5606},
  doi={10.1109/GLOBECOM48099.2022.10001144}}

@ARTICLE{tradeoff2,
  author={Du, Zhen and Liu, Fan and Xiong, Yifeng and Han, Tony Xiao and Eldar, Yonina C. and Jin, Shi},
  journal={IEEE Transactions on Signal Processing}, 
  title={Reshaping the ISAC Tradeoff Under OFDM Signaling: A Probabilistic Constellation Shaping Approach}, 
  year={2024},
  volume={72},
  number={},
  pages={4782-4797},
  doi={10.1109/TSP.2024.3465499}}

@ARTICLE{tradeoff3,
  author={Xiong, Yifeng and Liu, Fan and Cui, Yuanhao and Yuan, Weijie and Han, Tony Xiao and Caire, Giuseppe},
  journal={IEEE Transactions on Information Theory}, 
  title={On the Fundamental Tradeoff of Integrated Sensing and Communications Under Gaussian Channels}, 
  year={2023},
  volume={69},
  number={9},
  pages={5723-5751},
  doi={10.1109/TIT.2023.3284449}}

@ARTICLE{tradeoff4,
  author={Keskin, Musa Furkan and Mojahedian, Mohammad Mahdi and Lacruz, Jesus O. and Marcus, Carina and Eriksson, Olof and Giorgetti, Andrea and Widmer, Joerg and Wymeersch, Henk},
  journal={IEEE Transactions on Wireless Communications}, 
  title={Fundamental Trade-Offs in Monostatic ISAC: A Holistic Investigation Towards 6G}, 
  year={2025},
  volume={},
  number={},
  pages={1-16},
  doi={10.1109/TWC.2025.3563197}}

@misc{ericsson,
  author       = {Baldemair, Robert},
  title        = {Integrated Sensing and Communication},
  howpublished = {Ericsson Blog},
  year         = {2024},
  day          = {19},
  url          = {https://www.ericsson.com/en/blog/2024/6/integrated-sensing-and-communication},
  note         = {Accessed on 22 July 2025}
}

@misc{sivers,
  author       = {{Sivers Semiconductors}},
  title        = {{Evaluation Kit EVK06002 (57-71 GHz) - Sivers Semiconductors}},
  url          = {https://www.sivers-semiconductors.com/5g-millimeter-wave-mmwave-and-satcom/wireless-products/evaluation-kits/evaluation-kit-evk06002/},
  year         = {2021},
  note         = {Accessed July 22, 2025}
}




\newpage




\vfill

\end{document}